\documentclass[12pt,preprint]{aastex}

\shorttitle{Numerically determined transport laws for fingering convection}
\shortauthors{Traxler et al.}

\begin{document}


\title{Numerically determined transport laws for fingering (``thermohaline'') convection in astrophysics}


\author{A. Traxler$^{1}$ and P. Garaud$^{1,2}$}
\affil{$^1$ Department of Applied Mathematics and Statistics, Baskin
School of Engineering, UC Santa Cruz, 1156 High Street, Santa Cruz CA
95064 \\
$^2$ On sabbatical leave at: Institute for Astronomy, 34 `Ohi`a Ku St.,
Pukalani, HI 96768-8288}

\author{S. Stellmach$^3$}
\affil{$^3$ Institut f\"ur Geophysik, Westf\"alische Wilhelms-Universit\"at M\"unster, D-48149 M\"unster, Germany}




\begin{abstract}
We present the first three-dimensional simulations of fingering convection performed in a parameter regime close to the one relevant for astrophysics, and reveal the existence of simple asymptotic scaling laws for turbulent heat and compositional transport.  These laws can straightforwardly be extrapolated to the true astrophysical regime.  Our investigation also indicates that thermocompositional ``staircases,'' a key consequence of fingering convection in the ocean, cannot form spontaneously in the fingering regime in stellar interiors.  Our proposed empirically-determined transport laws thus provide simple prescriptions for mixing by fingering convection in a variety of astrophysical situations, and should, from here on, be used preferentially over older and less accurate parameterizations.  They also establish that fingering convection does not provide sufficient extra mixing to explain observed chemical abundances in RGB stars.
\end{abstract}



\section{Introduction}\label{sec:intro}

Recent years have seen rapidly growing interest in astrophysical applications
of fingering convection (often referred to as ``thermohaline
convection''), a process by which the unequal diffusion rates of thermal
and compositional fields can drive a double-diffusive instability and
lead to significant turbulent transport across stably-stratified
regions.  In the Earth's oceans this process has been well-studied as
``salt fingering'' (see \citealt{stern1960sfa}, or a recent review in \citealt{kunze2003ros}).  In stellar contexts, by contrast, its exact form and contribution to vertical mixing is much less clear.
Nevertheless, fingering convection has been invoked in various scenarios.
It is thought to explain the lack of observed metallicity signatures
from the infall of planets onto their host star \citep{vauclair2004mfa,garaud2010}.
It may also explain the presence of
extra mixing in the radiative zones of RGB stars needed to fit
abundance observations \citep{charbonnel2007tmp,stancliffe2010,denissenkov2010}.

While field, laboratory, and numerical measurements abound for salt fingering, data is much more scarce for the astrophysical regime.  Transport by fingering convection in astrophysics has until now been modeled through mixing-length theory, commonly used parameterizations being those of \citet{ulrich1972} and \citet{kippenhahn80}.
In these works, turbulent compositional transport depends sensitively on the finger aspect ratio.  As recently shown by \citet{denissenkov2010}, however, there is a fundamental contradiction between the large aspect ratio required to explain abundance observations of RGB stars using
these prescriptions \citep{charbonnel2007tmp}, and the near-unit aspect ratio of the ``fingers'' actually measured in his own two-dimensional simulations.

Three possibilities emerge to resolve this problem.  First, the
\citet{kippenhahn80} and \citet{ulrich1972} parameterizations may not adequately model
transport by fingering convection.  Secondly, two-dimensional
simulations may not correctly capture the true dynamics of this
inherently three-dimensional system.  Finally, thermocompositional
layers may form in the astrophysical objects studied, in which
case transport could be strongly enhanced above the level of
homogeneous fingering convection.  Indeed, ``staircases'' of
convectively mixed layers separated by thin fingering interfaces are
ubiquitous in the ocean.  In their presence, vertical mixing increases by as much as an order of magnitude above the already-enhanced mixing due to fingering \citep{schmitt2005edm,veronis2007}.  Whether such staircases form in astrophysics has never been established.

In this Letter we consider all three possibilities and present the first
three-dimensional simulations to address directly the question
of thermal and compositional transport by fingering convection in astrophysics.
Using a high-performance algorithm specifically designed for
homogeneous fingering convection, we are able to run simulations at moderately low
values of the Prandtl number (${\rm Pr}=\nu/\kappa_T$) and diffusivity ratio ($\tau=\kappa_\mu/\kappa_T$), where $\nu$ is viscosity, and $\kappa_T$ and $\kappa_\mu$ are the thermal and compositional diffusivities respectively.  We find, in Section \ref{subsec:smallflux}, asymptotic scaling laws for transport which are applicable to the stellar parameter regime (${\rm Pr}\ll 1, \tau\ll 1$). These provide a parameter-free,
empirically-motivated prescription for mixing by homogeneous fingering convection.
The compositional turbulent diffusivity derived is, as expected \citep{stern2001sfu} a few times larger than that obtained by \citet{denissenkov2010} for the same parameters, but remains too small to account for RGB abundance observations.

Since the presence of thermocompositional staircases is known to increase transport in the ocean, we next address the question of whether such staircases are likely to form spontaneously in the astrophysical context, in Section \ref{subsec:layers}.  Applying a
recently-validated theory for staircase formation \citep{radko2003mlf,stellmach2010},
we find that they
are in fact not expected to appear in this case.  We conclude in Section \ref{sec:conclusion} that our small-scale flux laws can reliably be used to estimate heat and compositional transport for a variety of astrophysical fingering systems and concur with Denissenkov's view that fingering convection is not sufficient to explain the required extra mixing in RGB stars.

\section{Model description and numerical algorithm}\label{sec:model}
The dynamics of double-diffusive systems depend on the fluid
properties ($\nu, \kappa_T, \kappa_\mu$), as well as its
local stratification, measured by $\nabla-\nabla_{\rm ad}$ (where
$\nabla=d\ln{T}/d\ln{p}$ and $\nabla_{\rm ad}=(d\ln{T}/d\ln{p})_{\rm
  ad}$ is the adiabatic gradient) and $\nabla_\mu$ (where
$\nabla_\mu=d\ln{\mu}/d\ln{p}$).
Here $T$ is temperature, $p$ is pressure, and $\mu$ is the mean molecular weight.
The fingering instability occurs in thermally stably-stratified systems ($\nabla-\nabla_{\rm ad} < 0$) destabilized by an adverse compositional gradient ($\nabla_\mu < 0$).
In most regimes of interest the main governing parameter is the density ratio, 
defined in astrophysics as \citep{ulrich1972}:
\begin{equation}
R_0^*=\frac{\nabla-\nabla_{\rm ad}}{\nabla_\mu}. \label{eq:density_ratio}
\end{equation}
This linear instability is well understood, thanks to early work in the oceanic context \citep{stern1960sfa,baines1969}, later extended in astrophysics by \citet{ulrich1972} and \citet{schmitt1983csf}.
A system is Ledoux-unstable when $R_0^*<1$, and fingering-unstable when $R_0^*\in [1,1/\tau]$.
Unfortunately, as discussed in Section \ref{sec:intro}, the nonlinear saturation of this instability, or in other words its turbulent properties, remain poorly understood.

We approach the problem from a numerical point of view, and model a fingering-unstable region using a local Cartesian frame $(x,y,z)$ with gravity ${\bf g} = -g{\bf e}_z$.
We run our numerical experiments using a high-performance spectral code, which was recently used to model three-dimensional fingering convection at oceanic parameters \citep{traxler2010}, as well as thermohaline staircase formation \citep{stellmach2010}.  As a consequence of its original purpose, our code uses
the Boussinesq approximation, which relates the density perturbations $\tilde{\rho}$ to the temperature and compositional perturbations $\tilde{T}$ and $\tilde{\mu}$ via
\begin{equation}
\frac{\tilde{\rho}}{\rho_0} = -\alpha \tilde{T} + \beta \tilde{\mu}, \label{eq:eos}
\end{equation}
where $\rho_0$ is a reference density, $\alpha= -\rho_0^{-1}\partial \rho/\partial T$, and $\beta= \rho_0^{-1}\partial \rho/\partial \mu$. 
Fingering convection is driven by constant large-scale temperature and compositional gradients $T_{0z}$ and $\mu_{0z}$.  These approximations, which greatly simplify the governing equations, are nevertheless justified since fingers are typically much smaller than a pressure or density scale height, and velocities are small compared with the sound speed.
As first shown by \citet{ulrich1972}, results in the Boussinesq case can straightforwardly be extended to the astrophysical case by replacing the Boussinesq density ratio $R_0=\alpha T_{0z}/\beta \mu_{0z}$ with
the true density ratio $R_0^*$ defined in equation (\ref{eq:density_ratio}).

We now describe the governing equations in more detail.
We express the velocity, temperature, and composition fields as the linear background stratification plus perturbations,
\begin{eqnarray}
T(x,y,z,t) & = & T_0(z) + \tilde{T}(x,y,z,t), \\
\mu(x,y,z,t) & = & \mu_0(z) + \tilde{\mu}(x,y,z,t), \\
{\bf u}(x,y,z,t) & = & {\bf \tilde{u}}(x,y,z,t),
\end{eqnarray}
where $T_0(z)=T_{0z}z$ and $\mu_0(z)=\mu_{0z}z$.  We nondimensionalize the system
noting that an appropriate length scale is provided by the anticipated finger width \citep{stern1960sfa}, $[l]=d=(\kappa_T\nu/g\alpha T_{0z})^{1/4}$.  We then scale time using $d^2/\kappa_T$, temperature with $T_{0z}d$, and composition with $(\alpha/\beta)T_{0z}d$.  The nondimensional governing equations are:
\begin{eqnarray}
\frac{1}{\mathrm{Pr}} \left(\frac{\partial{\bf \tilde{u}}}{\partial t} + {\bf \tilde{u}}\cdot\nabla {\bf \tilde{u}}\right) & = & -\nabla \tilde{p} + (\tilde{T} - \tilde{\mu}){\bf e}_z + \nabla^2 {\bf \tilde{u}} \label{eq:momentum}, \\
\nabla \cdot {\bf \tilde{u}} &=& 0 \label{eq:continuity},  \\
\frac{\partial \tilde{T}}{\partial t} + \tilde{w} + {\bf \tilde{u}}\cdot\nabla \tilde{T} & = & \nabla^2 \tilde{T} \label{eq:heat}, \\
\frac{\partial \tilde{\mu}}{\partial t} + \frac{1}{R_0}\tilde{w} + {\bf \tilde{u}}\cdot\nabla \tilde{\mu} & = & \tau \nabla^2 \tilde{\mu} \label{eq:composition},
\end{eqnarray}
Note that the Rayleigh number Ra then depends only on the computational domain height $L_z$:
\begin{equation}
{\rm Ra} = \frac{\alpha g T_{0z}L_z^4}{\kappa_T\nu} = \left(\frac{L_z}{d}\right)^4 \mbox{  . }
\end{equation}

We solve (\ref{eq:momentum}--\ref{eq:composition}) in a triply-periodic box of size $(L_x,L_y,L_z)$, so that
\begin{equation}
\tilde{T}(x,y,z,t) = \tilde{T}(x+L_x,y,z,t) = \tilde{T}(x,y+L_y,z,t) = \tilde{T}(x,y,z+L_z,t),
\end{equation}
and similarly for $\tilde{p}$, $\tilde{\mu}$ and $\tilde{\bf u}$.
\footnote{Because the length scale of the convective motions is set by the diffusive scales in the fingering problem, it does not suffer from the known pathology of triply-periodic thermal convection, {\em i.e.}\ the homogeneous Rayleigh-B\'{e}nard problem \citep{Calzavarinietal2006hrb}.}

\section{Results}\label{sec:results}
\subsection{Small-scale fluxes}\label{subsec:smallflux}

Six sets of simulations were conducted at successively lower values of
Pr and $\tau$, for a range of density ratio values listed in Table \ref{table_simsets} spanning in each case the instability range $R_0 \in [1,1/\tau]$.
As shown by \citet{traxler2010}, the statistical properties of fingering convection are independent of the computational domain size (i.e.\ of Ra and aspect ratio) provided it is wide enough to contain a sufficient number of fingers.  The domain must also be tall enough to ensure that individual fingers do not span the entire height of the box.
In most cases, the simulation domains were $67d\times 67d\times 107.2d$ with a resolution of $96^3$ grid points.  At the selected parameter values, this corresponds to approximately $5\times 5 \times 10$ wavelengths of the fastest-growing mode \citep[see][]{schmitt1979fgm},
thus allowing space for many fingers to fit in the domain.  As $R_0\rightarrow 1/\tau$, where fingers tend to become more vertically
elongated, wider and taller simulation domains ($83.75d\times 83.75d\times 268d$) were used to ensure that fingers remained uncorrelated across the box height.  In sets 5 and 6 (the two lowest-diffusivity runs), lower finger heights allowed for shorter computational boxes ($67d\times 67d\times 67d$) but a higher resolution ($192^3$ grid points) was used to resolve the finer-scale structures.
Near $R_0=1$, where the system dynamics are most turbulent, one additional simulation with twice the vertical resolution (192 or 384 points vertically) was run for each set.  In all cases, the measured fluxes in the more fully-resolved run differed by no more than a few percent from the lower-resolution run, confirming that the spatial resolution used was sufficient to extract accurate fluxes.  Sample snapshots of two of our simulations are shown in Figure \ref{finger_renders}.
As an important note, we also find that, except when $R_0\rightarrow 1/\tau$, the finger aspect ratio is close to unity \citep{denissenkov2010}.

\begin{figure}
\begin{center}
\epsscale{0.5}
\plotone{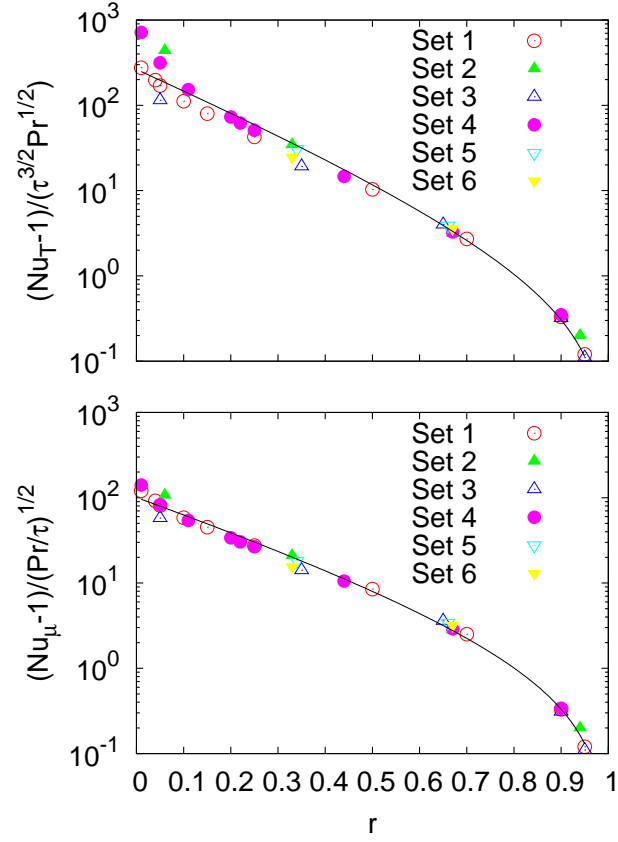}
\caption{Turbulent heat and compositional fluxes as a function of
  rescaled density ratio $r$.  Dividing the turbulent heat flux by $\tau^{3/2}{\rm Pr}^{1/2}$ and compositional flux by $\sqrt{{\rm Pr}/{\tau}}$ reveals the universal scaling laws (\ref{Nuscale}) and (\ref{NuCscale}).  Also shown are $f(r)$ and $g(r)$ fitted to the data (solid lines).
  \label{fig_Nu_fits}}
\end{center}
\end{figure}

We define non-dimensional heat and compositional fluxes through
the Nusselt
numbers ${\rm Nu}_T$ and ${\rm Nu}_\mu$, ratios of the total flux to the diffusive flux of the field considered.  Equivalently, ${\rm Nu}_{T,\mu}-1$ represents the ratio of the turbulent diffusivity to the microscopic diffusivity.
We measure ${\rm Nu}_T$ and ${\rm Nu}_\mu$ from each simulation once the system has settled into a statistically-steady state. 
The results are summarized in Figure \ref{fig_Nu_fits}.

For ease of comparison between different simulation
sets, we defined the rescaled density ratio as $r=(R_0-1)/(\tau^{-1}-1)$ so that the instability range
is $r \in [0,1]$ in all cases. We find that {\it all simulation sets collapse onto a single universal profile} for the turbulent fluxes $\mathrm{Nu}_T(r)-1$ and $\mathrm{Nu}_\mu(r)-1$ as Pr and $\tau$ decrease, profiles which can be expressed as
\begin{eqnarray}
\mathrm{Nu}_T(r)-1 & = & \tau^{3/2}\sqrt{\mathrm{Pr}} f(r), \label{Nuscale} \\
\mathrm{Nu}_\mu(r)-1 & = & \sqrt{\frac{\mathrm{Pr}}{\tau}}g(r), \label{NuCscale}
\end{eqnarray}
where the universal functions $f(r)$ and $g(r)$ are adequately fitted to the data using
\begin{equation}
g(r), f(r) \sim ae^{-br}(1-r)^{c} \label{Nufit}.
\end{equation}
For $f(r)$ the data suggests that $a=264$, $b=4.7$, $c=1.1$, and for $g(r)$ we find $a=101$, $b=3.6$, $c=1.1$.
Figure \ref{fig_Nu_fits} shows these fits together with the rescaled Nusselt numbers.
The physical interpretation of these scalings with $\tau$ and Pr is the following:  for $\nu, \kappa_\mu \ll \kappa_T$, the instability is driven by the compositional field and the dependence of compositional transport on $\kappa_T$ must drop out. The turbulent compositional diffusivity $D_\mu=({\rm Nu}_\mu-1)\kappa_\mu$ is then proportional to the geometric mean of $\nu$ and $\kappa_\mu$. The scaling for the temperature Nusselt number then follows by assuming that the nondimensional turbulent flux ratio $\gamma=<\tilde{w}\tilde{T}>/<\tilde{w}\tilde{\mu}>$ is always $O(1)$ \citep{radko2003mlf}.

Although these simulations were conducted at parameter values a few orders of magnitude higher than those appropriate for stellar interiors, we find that they follow universal asymptotic laws. The simplicity of the system
considered (e.g. no boundaries) also suggests that these laws {\it can} reliably be extrapolated to the astrophysical regime (as long as Pr is of order $\tau$).
From this we draw two critical observations.
As found by \citet{denissenkov2010}, the contribution of fingering convection to heat transport at stellar parameter values (${\rm Pr}\sim 10^{-6}, \tau\sim 10^{-6}$) is entirely negligible.  It therefore does not affect the thermal structure of the object in any way.  Compositional transport, on the other hand, can be significantly enhanced above molecular diffusion, by about two orders of magnitude depending on the ratio ${\rm Pr}/\tau=\nu/\kappa_\mu$.

\begin{figure}
\begin{center}
\plotone{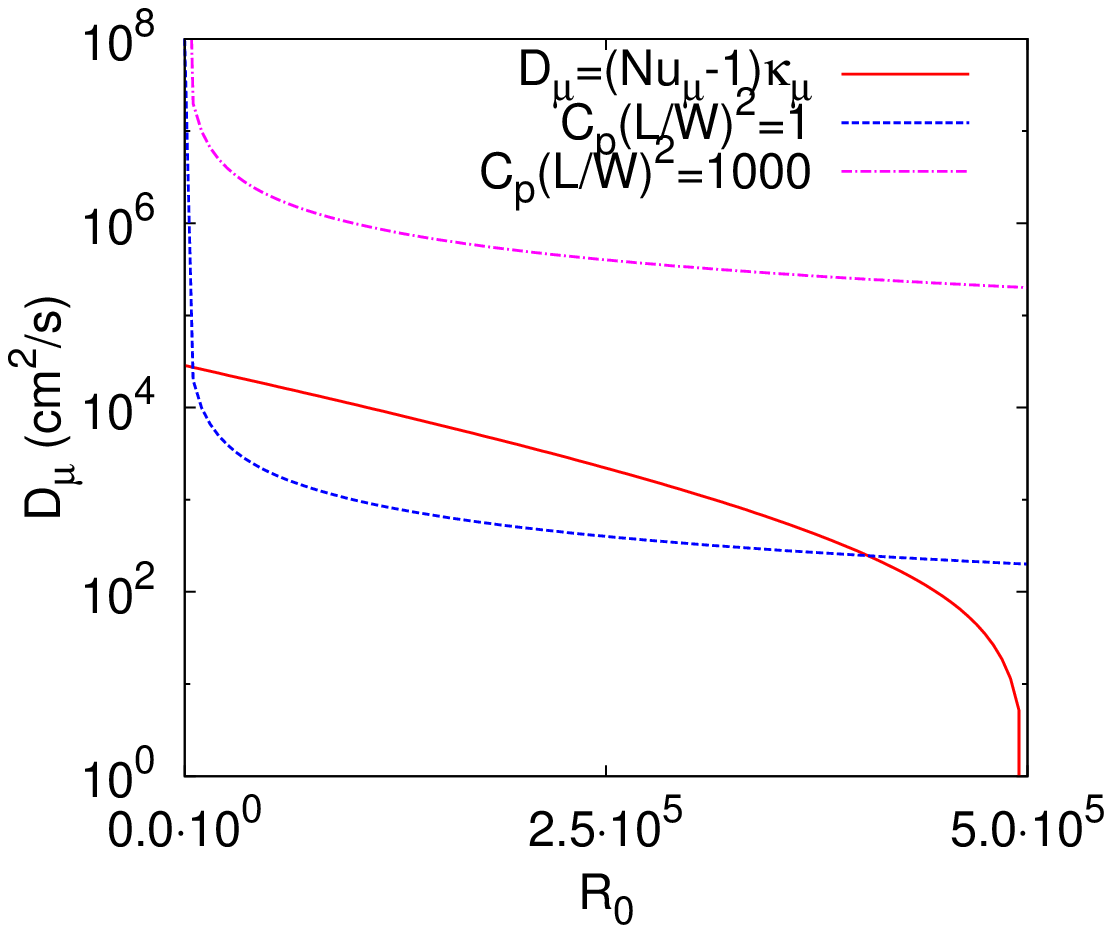}
\caption{Comparison between our empirically-determined compositional turbulent diffusivity and the parameterizations of \citet{kippenhahn80} and \citet{ulrich1972} (where the coefficient $C_p(L/W)^2$ serves as a free parameter).  Since we find $L/W\sim 1$, \citet{kippenhahn80} underestimates $D_\mu$ for most of the $R_0$ range and overestimates it as $R_0\rightarrow 1$ and $R_0\rightarrow 1/\tau$. \citet{ulrich1972} fares slightly better at intermediate $R_0$.  Our results show that turbulent mixing by fingering convection is not sufficient to explain observations, for which $C_p(L/W)^2 \sim 1000$ is required (see text for detail).
\label{fig:mixing}}
\end{center}
\end{figure}

We now compare our flux laws with the parameterization of
\citet{ulrich1972} and \citet{kippenhahn80}, in which ${\rm Nu}_\mu-1 = C_p (L/W)^2/\tau R_0^*$, where $L/W$ is the aspect ratio of the fingers (length $L$ divided by width $W$) and $C_p$ is a model constant ($C_p=1$ for \citealt{kippenhahn80}, $C_p=8\pi^2/3\approx 26$ for \citealt{ulrich1972}).  Figure \ref{fig:mixing} compares $D_\mu$ as a function of density ratio, at the fluid parameters used by \citet{denissenkov2010}, ${\rm Pr}=4\times 10^{-6}$ and $\tau=2\times 10^{-6}$.
Given that the numerically-determined finger aspect ratio is close to one for all simulations as $\tau\ll 1$, ${\rm Pr}\ll 1$, we find that the \citet{ulrich1972} prescription fares comparatively better than \citet{kippenhahn80}.  However, it still vastly
overestimates mixing both near $R_0=1$ (where the system is closest to Ledoux instability) and $R_0=1/\tau$, close to linear stability.
Crucially, our results show that the diffusivity obtained is not sufficient to explain RGB abundance observations for which a value of $C_p(L/W)^2\approx 1000$ is needed \citep{charbonnel2007tmp,denissenkov2010}.

\subsection{Layer formation}\label{subsec:layers}

The small-scale flux laws presented above are accurate measurements of transport by {\it homogeneous} fingering convection.
However, they may not appropriately
describe transport in the presence of the kind of large-scale
thermocompositional staircases which are known to form in oceanic thermohaline
convection \citep[{\em e.g.}\ ][]{schmitt2005edm}.  The origin of such staircases
has recently been discovered by \citet{radko2003mlf}, who demonstrated the existence of
a positive feedback mechanism that can, in some circumstances, drive the growth of
horizontally-invariant perturbations in the temperature and salinity fields:
large-scale variations of the density ratio lead to convergences and divergences of fingering fluxes that reinforce the original perturbation, producing a growing disturbance that ultimately overturns into regular ``steps.''  This mechanism relies fundamentally on the variation of the flux ratio $\gamma=F_T/F_S$ (where $F_T$ is the turbulent heat flux and $F_S$ is the turbulent salt flux) with density ratio, and was therefore called the $\gamma$-instability.
The $\gamma$-instability mechanism accurately predicts the growth rate of large-scale perturbations, and subsequent overturning into layers, in both two-dimensional simulations \citep{radko2003mlf},
and three-dimensional simulations of staircase formation \citep{traxler2010,stellmach2010}.
The instability requires that $\gamma$ decreases as $R_0$ increases, a condition which must be experimentally determined from the fluxes for each set of fluid parameters Pr and $\tau$.

In Radko's original formulation for oceanic staircases, the systems of interest were dominated by turbulent fluxes. Diffusive contributions to the total fluxes were neglected for convenience and simplicity.  However, as shown in Section \ref{subsec:smallflux}, turbulent heat transport is negligible in the astrophysical case, so the $\gamma$-instability theory is re-derived here including all diffusive terms for accuracy and completeness.  We begin with the equations of Section \ref{sec:model} and average them over many fingers:
\begin{eqnarray}
\frac{1}{\mathrm{Pr}} \left(\frac{\partial\mathbf{u}}{\partial t} + \mathbf{u}\cdot\nabla \mathbf{u} \right) & = & -\nabla p + (T - \mu){\bf e}_z + \nabla^2 \mathbf{u} - \frac{1}{\mathrm{Pr}}\nabla \cdot \mathbf{R} \label{eq:mean_momentum}, \\
\frac{\partial T}{\partial t} + w + \mathbf{u}\cdot\nabla {T}
& = & \nabla^2 T - \nabla \cdot
\mathbf{F}_T, \label{eq:mean_temperature} \\
\frac{\partial \mu}{\partial t} + \frac{1}{R_0}w + {\bf u}\cdot\nabla \mu & = & \tau \nabla^2 \mu - \nabla \cdot \mathbf{F}_\mu \label{eq:mean_composition},
\end{eqnarray}
where $\mathbf{u}$, $T$, and $\mu$ now represent averaged large-scale
fields.  On the right-hand sides appear the usual Reynolds stress term,
$R_{ij}=\overline{\mathbf{\tilde{u}}_i\mathbf{\tilde{u}}_j}$, and the turbulent
fluxes $\mathbf{F}_T=\overline{\mathbf{\tilde{u}}\tilde{T}}$,
$\mathbf{F}_\mu=\overline{\mathbf{\tilde{u}}\tilde{\mu}}$.

We proceed as in earlier formulations \citep{radko2003mlf,traxler2010} by making
several simplifying assumptions, the first being
that Reynolds stresses are small enough to neglect and the second
being that the turbulent fluxes are dominated by their vertical
component, so that $\mathbf{F}_T\approx F_T \mathbf{e}_z$,
$\mathbf{F}_\mu \approx F_\mu \mathbf{e}_z$.  
The $\gamma$-instability involves only the temperature and compositional fields,
so we consider only zero-velocity perturbations and discard the momentum equation.
We then define the total fluxes as well as their ratio:
\begin{eqnarray}
F_T^{\rm tot} & = & F_T - (1 + \partial T/\partial z), \\
F_\mu^{\rm tot} & = & F_\mu - \tau\, (1/R_0 + \partial \mu/\partial z), \\
\gamma^{\rm tot} & = & \frac{F_T^{\rm tot}}{F_\mu^{\rm tot}}.
\end{eqnarray}
so that the thermal and compositional Nusselt numbers are:
\[
{\rm Nu}_T = \frac{F_T^{\rm tot}}{-(1+\partial T/\partial z)}\ ,\ {\rm Nu}_\mu = \frac{F_\mu^{\rm tot}}{-\tau(R_0^{-1}+\partial \mu/\partial z)}.
\]
The final assumption is that ${\rm Nu}_T$, ${\rm Nu}_\mu$, and $\gamma^{\rm tot}$ depend
only on the local value of the density ratio $R_\rho$.  Note that $R_\rho \ne R_0$ and may vary with $z$.
We now linearize equations (\ref{eq:mean_temperature}) and (\ref{eq:mean_composition}) around a state of homogeneous turbulent convection in which $T = 0 + T'$, $\mu = 0 + \mu'$, and $R_\rho = R_0 + R'$.
For example, linearizing the density ratio $R_\rho$, we have:
\begin{eqnarray}
R_\rho & = & \frac{\alpha T_{0z}(1 + \partial T'/\partial z)}{\beta \mu_{0z}[1 + (\frac{\alpha T_{0z}}{\beta \mu_{0z}}) \partial \mu'/\partial z]}, \nonumber \\
 & \approx & R_0(1 + \partial T'/\partial z - R_0 \partial \mu'/\partial z). \label{R0_linear}
\end{eqnarray}
The temperature equation yields
\begin{eqnarray}
\frac{\partial T'}{\partial t} & = & -\frac{\partial F_T^{\rm tot}}{\partial z}, \nonumber \\
 & = & \frac{\partial {\rm Nu}_T}{\partial z} + {\rm Nu}_T\frac{\partial^2 T'}{\partial z^2}, \nonumber \\
 & = & \left. \frac{\partial {\rm Nu}_T}{\partial R_\rho}\right|_{R_0} \frac{\partial R_\rho}{\partial z} + {\rm Nu}_T(R_0)\, \frac{\partial^2 T'}{\partial z^2}, \nonumber \\
 & = & \left. \frac{\partial {\rm Nu}_T}{\partial R_\rho}\right|_{R_0} R_0 \left(\frac{\partial^2 T'}{\partial z^2} - R_0 \frac{\partial^2 \mu'}{\partial z^2}\right) + {\rm Nu}_T(R_0)\, \frac{\partial^2 T'}{\partial z^2}.  \label{T_linear}
\end{eqnarray}
and similarly for equation (\ref{eq:mean_composition}).  
Assuming normal modes of the form $T', \mu' \sim e^{ikz + \lambda t}$,
equations (\ref{T_linear}) and the equivalent linearization of (\ref{eq:mean_composition}) combine into the quadratic
\begin{equation}
\lambda^2 + \lambda k^2 \left[{\rm Nu}_{0}(1-A_1R_0) +
  A_2\left(1-\frac{R_0}{\gamma_0^{\rm tot}}\right)\right] - k^4 A_1{\rm
  Nu}_{0}^2 R_0 = 0, \label{eq:quadratic}
\end{equation}
where we use the following notation for simplicity:
\[
\begin{array}{ll}
A_1 = R_0 \left. \frac{d(1/\gamma^{\rm tot})}{dR_\rho} \right|_{R_0}, & A_2 = \left. R_0 \frac{d {\rm Nu}_T}{dR_\rho} \right|_{R_0}, \\
{\rm Nu}_{0} = {\rm Nu}_T(R_0), & \gamma_0^{\rm tot} = \gamma^{\rm tot}(R_0).
\end{array}
\]
Note that this is identical to the quadratic of \citet{radko2003mlf} replacing $\gamma$ by $\gamma^{\rm tot}$.

\begin{figure}
\begin{center}
\plotone{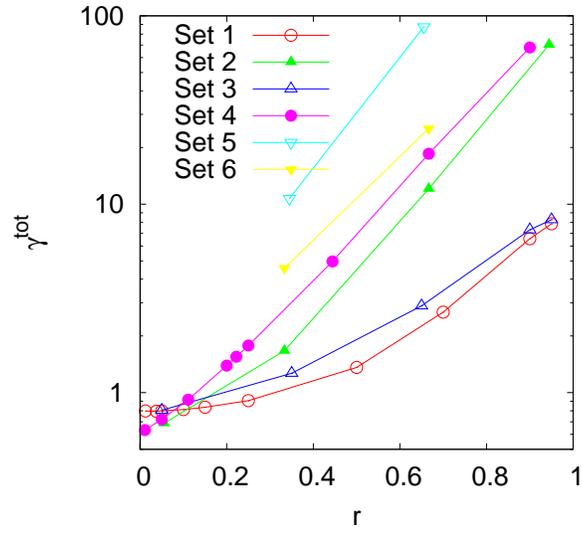}
\caption{The total flux ratio $\gamma^{\rm tot}=F_T^{\rm tot}/F_\mu^{\rm tot}$ for various systems.  As the density stratification becomes more stable ($r\rightarrow 1$), the fluxes of heat and composition are dominated by diffusive contributions and their ratio approaches the limiting value $R_0/\tau$. The
flux ratio increases monotonically with $r$, indicating that staircases will not form spontaneously in this parameter regime (${\rm Pr}\ll 1, \tau \ll 1$).\label{fig:gammatot}}
\end{center}
\end{figure}

As discussed by \citet{radko2003mlf}, a sufficient condition for the instability is that $\gamma^{\rm tot}$ is a decreasing function of density ratio.
Indeed, when this is the case, the coefficient $A_1$ is positive, and the expression (\ref{eq:quadratic}) has two real roots, one positive and one negative.
In order to determine whether the $\gamma$-instability occurs and
therefore whether staircases may spontaneously form at low Pr, low $\tau$, we simply need to calculate $\gamma^{\rm tot}$ from our previous measurements.  The results are
shown in Figure \ref{fig:gammatot}.  We find that the flux ratio always increases with density ratio. Crucially, this suggests that thermocompositional layers are not expected to form spontaneously in astrophysical fingering convection.  In the absence of such ``staircases,'' the fluxes are accurately supplied by our flux laws \ref{Nuscale} and \ref{NuCscale}.

\section{Conclusion}\label{sec:conclusion}

Our results can be summarized in a few key points.
Using high-performance three-dimensional numerical simulations we find that turbulent compositional transport by fingering convection follows the simple law:
\begin{equation}
D_\mu = \kappa_\mu ({\rm Nu}_\mu-1) = 101 \sqrt{\kappa_\mu \nu} e^{-3.6r}(1-r)^{1.1},
\end{equation}
where $r=(R_0^*-1)/(\tau^{-1}-1)$.
Meanwhile, the turbulent heat flux is negligibly small at stellar parameters \citep[see][]{denissenkov2010}.
We also find that
large-scale thermocompositional staircases are not expected to form spontaneously for Pr , $\tau \ll 1$, so that the turbulent diffusivity proposed above may and should be used as given to
model astrophysical fingering convection.

Applying these findings to the problem of RGB stars, we concur with Denissenkov's conclusion that mixing by fingering convection alone cannot explain the observed RGB abundances, and that additional mechanisms should be investigated instead \citep[{\em e.g.}\ gyroscopic pumping, see][]{garaud2010gyr}.  A second example of application to the planetary pollution problem is presented in a companion paper \citep{garaud2010}.

\acknowledgments

A.T. and P.G. are supported by the National Science Foundation, NSF-0933759. S.S. was supported by grants from the NASA Solar and Heliospheric Program (NNG05GG69G, NNG06GD44G, NNX07A2749).

\begin{figure}
\begin{center}
\plottwo{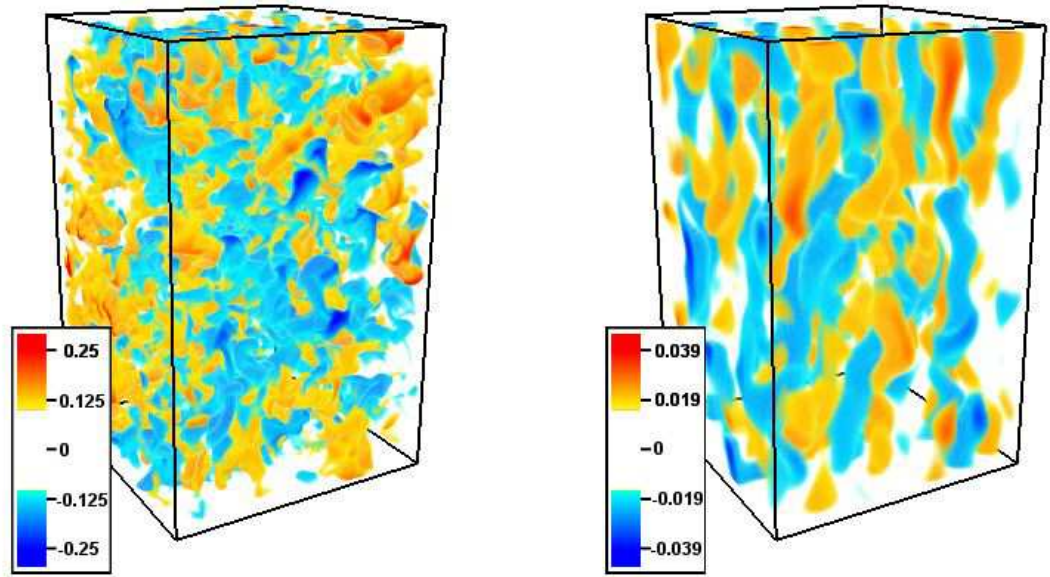}{fig1b.epsf}
\caption{Compositional perturbations, in units of the compositional contrast across the height of the domain, of two simulations at ${\rm Pr}=\tau=1/10$, at density ratio $R_0=1.45$ (left) and $R_0=9.1$ (right).  In general, fingering convection is more laminar, and the fingers are taller as $R_0\rightarrow 1/\tau$ (as the background stratification becomes more stable).  However, as Pr and $\tau$ decrease, secondary instabilities enforce shorter fingers even at high density ratio.\label{finger_renders}}
\end{center}
\end{figure}

Figure \ref{finger_renders} was rendered using VAPOR \citep{vapor1,vapor2}, a product of the National Center for Atmospheric Research.

\clearpage

\clearpage

\begin{table}
\begin{center}
\caption{Summary of governing parameters for all simulation sets.\label{table_simsets}}
\begin{tabular}{clcl}
\tableline\tableline
Set   & Pr & $\tau$ & $R_0$ \\ \tableline
1 & 1/3  & 1/3    & $1.025, 1.075, 1.1, 1.1^{\alpha}, 1.1^{\beta},$ \\
 & & & $1.125, 1.2, 1.3, 1.5, 2, 2.4,$ \\
 & & & $2.8, 2.9^{a,\alpha}$ \\
2 & 1/3  & 1/10   & $1.5, 1.5^{\beta}, 4, 7, 9.5^{a,\alpha}$ \\
3 & 1/10    & 1/3    & $1.1, 1.1^{\alpha}, 1.7, 2.3, 2.8^{a,\alpha},$ \\
 & & & $2.9^{a,\alpha}$ \\
4 & 1/10    & 1/10   & $1.1, 1.45, 1.45^{\alpha}, 2, 2.8, 3,$ \\
 & & & $3.3, 5, 7, 9.1^{a,\alpha}$ \\
5 & 1/10    & 1/30   & $11^{b,\gamma}, 11^{b,\delta}, 20^{b,\gamma}$ \\
6 & 1/30 & 1/10   & $4^{b,\gamma}, 7^{b,\gamma}$ \\
\tableline
\end{tabular}
\tablecomments{Relevant parameters include values of Pr and $\tau$, of the density ratio $R_0\in [1,1/\tau]$, domain size, and resolution.  Unless otherwise specified (see footnotes), simulations were run at domain sizes of $67\times 67\times 107.2d$ and resolution of $96^3$ grid points.}
\tablenotetext{a}{$83.75d \times 83.75d\times 268d$}
\tablenotetext{b}{$67d\times 67d\times 67d$}
\tablenotetext{\alpha}{$96\times 96\times 192$}
\tablenotetext{\beta}{$192\times 192\times 384$}
\tablenotetext{\gamma}{$192\times 192\times 192$}
\tablenotetext{\delta}{$384\times 384\times 384$}
\end{center}
\end{table}

\end{document}